\documentclass{cernrep} 
\usepackage{texnames}
\usepackage[T1]{fontenc}
\usepackage{amsmath}
\usepackage{accents}
\newlength{\dhatheight}
\newcommand{\doublehat}[1]{%
    \settoheight{\dhatheight}{\ensuremath{\hat{#1}}}%
    \addtolength{\dhatheight}{-0.35ex}%
    \hat{\vphantom{\rule{1pt}{\dhatheight}}%
    \smash{\hat{#1}}}} 

\begin{document}
\title{RooStats for Searches}
 
\author{Gr$\acute{\textrm{e}}$gory Schott, on behalf of the RooStats team}

\institute{KIT, Institut f\"ur Experimentelle Kernphysik, Karlsruhe, Germany}

\setcounter{page}{199}
\maketitle
 
\begin{abstract}
The RooStats toolkit, which is distributed with the ROOT software package,
provides a large collection of software tools that implement statistical methods
commonly used by the
High Energy Physics community. The toolkit is based on RooFit, 
a high-level data analysis modeling package that
implements various methods of statistical data analysis. 
RooStats enforces a clear mapping of statistical concepts to C++ classes
and methods and emphasizes the
ability to easily combine analyses within and across experiments. We present an overview of the RooStats toolkit,
describe some of the methods used for hypothesis testing and
estimation of confidence intervals and finally discuss some of the
latest developments.
\end{abstract}
 
\section{Introduction}

The RooStats project~\cite{RooStatsProceeding,RooStatsWebpage} is a
collaborative open source project initiated by members of ATLAS, CMS and
the CERN ROOT team. The RooStats toolkit --- based on previously existing code used in ATLAS~\cite{kyle} and 
CMS~\cite{Piparo}, which has been extended and
improved --- has been distributed with ROOT since summer
2008.  The toolkit provides and consolidates statistical tools needed for LHC
analyses and allows one to apply and compare the most popular and
well-established statistical approaches. Thanks to readily available well-known
tools, results across experiments can be better understood and
compared. This is not only a desirable feature but also a required one
when it comes to combining analysis results as will be discussed
later. Finally, the RooStats project aims to provide reasonably flexible, well-tested, 
documented tools. The RooStats developments benefit from
scientific oversight from the statistics committees of both
experiments.

In High Energy Physics, the goal of an analysis is usually to
test a prediction or search for new physics, leading to the 
estimation of the statistical significance of a possible observation
or the construction of confidence intervals --- often expressed as upper
or lower limits in case of a non-observation. The most common
statistical procedures are:
\begin{itemize}
\item point estimation: \ie the determination of the best estimate of
parameters of the model,
\item confidence or credible interval estimation: \ie regions
representing the range of parameters of interest compatible with the
data,
\item hypothesis tests: 
\ie comparing the data to two or more hypotheses,
\item goodness of fit: to quantify how well a given model describes the
observed data.
\end{itemize} 
RooStats aims to cover some of these common statistical procedures.

The RooStats package is built on top of RooFit~\cite{RooFit}, which is
a data modeling toolkit developed originally within the BaBar
collaboration and now integrated into ROOT. The most crucial
element of RooFit is its ability to model probability densities,
likelihood functions, and data, in a very flexible way that 
can deal with arbitrarily complex cases. Some recent developments in
RooFit provide additional tools specifically needed
by RooStats. The RooStats code is organized into three groups of classes: 
\emph{calculators} that perform the
statistical calculations, \emph{results} and \emph{utilities}
that facilitate the RooStats work flow.

After a few generalities, given in Sect.~\ref{sec:generalities}, the
classes implementing statistical inferences and results 
are discussed in Sect.~\ref{sec:RooStats}. In
Sect.~\ref{sec:utils}, we describe RooStats utilities, while  
Sect.~\ref{sec:theend} will have a few words on some
applications and perspectives.

\section{Generalities}
\label{sec:generalities}

We begin by clarifying some of the terminology commonly used:
\begin{itemize}
\item \emph{Observables}: quantities that are
measured by an experiment (\eg mass, helicity angle, output
of a neural network) that form a \emph{data set}.
\item \emph{Model}: the probability density function (PDF) --- either parametric
or non-parameteric --- that
describes one or multiple observables and normalized so that their integral over any
observable is unity.
\item \emph{Parameters of interest}: parameters of the model whose value we wish
to estimate or constrain (\eg a particle mass or a cross-section).
\item \emph{Nuisance parameters}: uncertain parameters of the model other than the ones of interest (\eg parameters associated with systematics, such as normalization or shape
parameters). The treatment of nuisance parameters varies according to the 
statistical approach.
\end{itemize}

\subsection{Likelihood Function}

The modeling of  the likelihood function is the principal task of RooFit.
RooFit, which builds on ROOT, maps
mathematical concepts to RooFit classes. For example, variables,
functions, probability densities, integrals, a space point, or a list thereof, are
handled by {\tt RooRealVar}, {\tt RooAbsReal}, {\tt
RooAbsPdf}, {\tt RooRealIntegral}, {\tt RooArgSet} and {\tt
RooAbsData}, respectively. A large collection of functions are available to describe
the PDF. The functions are handled by classes inheriting from {\tt
RooAbsPdf} and can be easily combined to build arbitrarily complex
models through addition, multiplication, and convolution. 
For both data and models there exist some binned and
unbinned representations. For each model, integration and maximum
likelihood fitting is supported and utilities are provided for the Monte Carlo
generation of pseudo data, in order to perform  "toy" studies, and for the visual inspection 
of results. The 
utilities and great modularity of RooFit are the principal factors that  drove the choice
of RooFit as the basis of RooStats. One can work with arbitrarily
complex data and models 
and one can handle large sets of observables and parameters.

Most statistical methods usually start with a likelihood function. A
rather general likelihood function, for use in our field, with multiple
observables, can be written as:
\begin{equation}
  \label{eqn:one} L(\mathbf{x}|r, s, b, \mathbf{\theta_s},
\mathbf{\theta_b}) = e^{-(rs + b)} \prod^{n}_{j=1}
[r s f_{s}(\mathbf{x_j}|\mathbf{\theta_s})+bf_{b}(\mathbf{x_j}|\mathbf{\theta_b})].
\end{equation}
The PDFs $f_s$ and $f_b$ represent the distributions of observables $\mathbf{x}$ for the signal
and background, with
parameters $\mathbf{\theta_s}$ and $\mathbf{\theta_b}$, respectively. The
parameters $s$ and $b$ --- typically, the expected signal and background counts, respectively --- are
constrained by the number $n$ of observed events\footnote{Sometimes described
as an extended likelihood; it can also be viewed as the limit of a binned multi-Poisson likelihood function with arbitrarily small bins.}.  In this likelihood function a strength factor $r$
multiplies the expected number of signal events\footnote{This is sometimes
done to redefine the parameter of interest such that $r$ is the ratio of the
signal production cross-section to the expected value of the
cross-section. For example, in the search for the Standard
Model Higgs boson, obtaining a $95\%$ CL upper-limit for $r=1$ means
the Standard Model Higgs hypothesis can be excluded at $95\%$ CL.}.

\subsection{Model Configuration}

Before one can perform a statistical inference, it is necessary to specify
the model: the PDF of possible observables, the actual observables, the
parameters of interest, the nuisance parameters, possibly a Bayesian
prior, \etc The RooStats calculators can be configured, via the constructor, either with 
the model specifications given as individual
RooFit objects or with a  {\tt ModelConfig} object,
in which the model specification
is bundled. For most
of the calculators both configuration mechanisms are available. 
The idea behind {\tt ModelConfig} is to provide a uniform way
to configure calculators. The downside is that it
becomes less obvious what elements of the {\tt ModelConfig} are
necessary for a given calculator. For example, the prior probability
will not be used in frequentist-based calculations while the list
of observables, which is mainly used to generate pseudo-data, is not needed when
computing Bayesian limits.

The model is often completed by a set of observed data. Moreover, 
the calculators can be configured for a number of options
specific to the statistical algorithms (\eg number of Monte Carlo
iterations, size of the test, test statistic, \etc). Finally, the
calculator is run and returns the result of a hypothesis test or a confidence
interval.

\section{RooStats Calculators}
\label{sec:RooStats}

Below, we describe the RooStats calculators, which are based on the following  conceptual approaches:
\begin{itemize}
\item \emph{Classical or Frequentist}: this school of statistics restricts
itself to statements of the form "probability of the data given
the hypothesis". Probability is interpreted as a limit of 
relative frequencies of various outcomes.
\item \emph{Bayesian}: this school of statistics views probability more
broadly, which permits 
statements of the form "probability of the hypothesis given the data".
Typically, probability is interpreted as a "degree of belief" in the veracity
of an hypothesis.
\item \emph{Likelihood}: this approach uses a
frequentist notion of probability (\eg it does not require the specification of a prior for
the hypothesis), but inferences are not guaranteed to satisfy some
frequentist properties
(\eg coverage). Like the Bayesian approach, this likelihood approach  obeys the likelihood principle, while frequentist methods do not.
\end{itemize} 
We give a brief description of the methods available in
RooStats and refer the reader to textbook literature
for details (see, for example~\cite{James,PDG}).

\begin{figure}
\centering\includegraphics[width=1\linewidth]{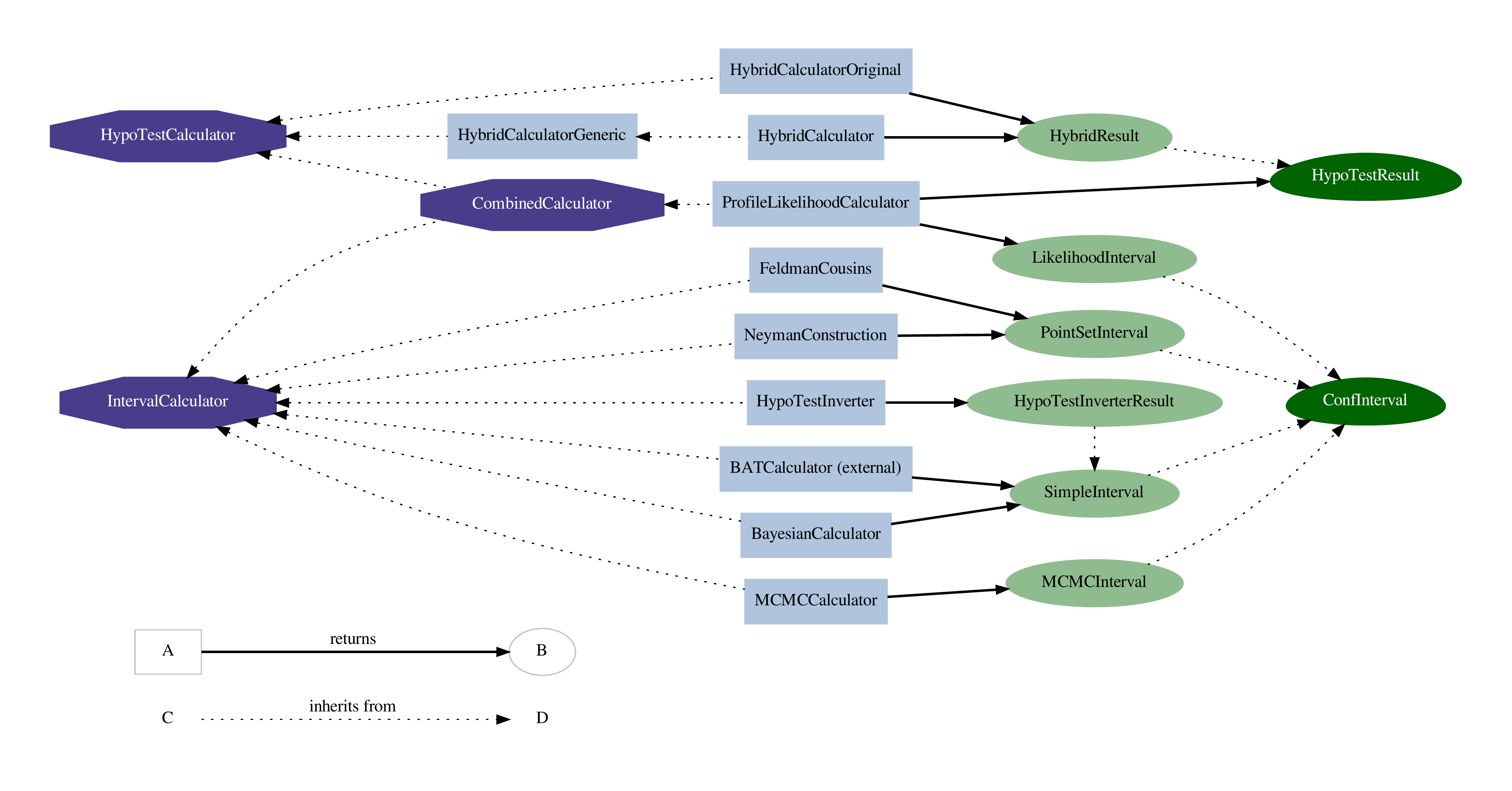}
\caption{ Diagram of the interfaces for hypothesis testing and
confidence interval calculations and classes used to return the
results of these statistical tests. \label{fig:first} }
\end{figure}

As can be seen from Fig.~\ref{fig:first}, there are two general
classes of calculators in RooStats: those performing hypothesis-tests
and those computing confidence or credible intervals, which inherit,
respectively, from the classes {\tt HypoTestCalculator} and {\tt
IntervalCalculator} and return, respectively, objects inheriting from
the classes {\tt HypoTestResult} or {\tt ConfInterval}.

The {\tt IntervalCalculator} interface allows the user to provide the
model, the data set, the parameters of interest, the nuisance
parameters and the size $\alpha$ of the test ($\alpha=1-\mathrm{CL}$,
where $\mathrm{CL}$ is the confidence/credible level). After
configuring the calculator, a {\tt ConfInterval} pointer is returned
via the method {\tt IntervalCalculator::GetInterval()}. Depending on
the calculator used, a different type of {\tt ConfInterval} will be
returned (\eg connected interval, multi-dimensional interval, \etc) but
each shares the ability to test if a point lies within the
interval using the method {\tt ConfInterval::} {\tt IsInInterval(p)}.

The {\tt HypoTestCalculator} can be configured with the model, the
data and parameter sets specifying the two hypotheses to be tested. Through
{\tt HypoTestCalculator::GetHypoTest()}, a pointer to the result can
be retrieved and the result object can be queried for $p$-values and the corresponding
significances, or $Z$-values, found by equating a $p$-value to a one-sided Gaussian tail probability
and solving for the number of standard
deviations. In this convention, a $p$-value of $2.87\times10^{-7}$
corresponds to a $Z$-value of $5\sigma$.

\subsection{Profile-Likelihood Calculator}

The {\tt ProfileLikelihoodCalculator} class implements a
likelihood-based method to estimate a confidence interval and to perform
an hypothesis test for a given parameter value. To illustrate the
method, let us assume that the likelihood function depends on a set
$K$ parameters $\mathbf\theta$, one of which is the parameter of
interest. From the likelihood function
$L(\mathbf{x}|\theta_0,\mathbf{\theta}_{i \neq 0})$, similar to the one
of Eq.~(\ref{eqn:one}) but where the parameter of interest $r$ has
been renamed $\theta_0$, for generality, the profile likelihood
function is the numerator in the ratio: 
\begin{equation}
    \label{eqn:profile_likelihood} \lambda(\theta_{0} ) = \frac{
L(\theta_{0},\doublehat{\mathbf{\theta} }_{i \neq 0} ) } {
L(\hat{\theta}_{0}, \hat{\mathbf{\theta}}_{i \neq 0} ) }.
\end{equation} 
The denominator, $L(\hat{\theta})$ is the absolute maximum of the
likelihood, while the
numerator is the maximum value of the likelihood for a 
given value of $\theta_{0}$.

\begin{figure}[hbt]
  \begin{center}
    \includegraphics[width=0.75\textwidth]{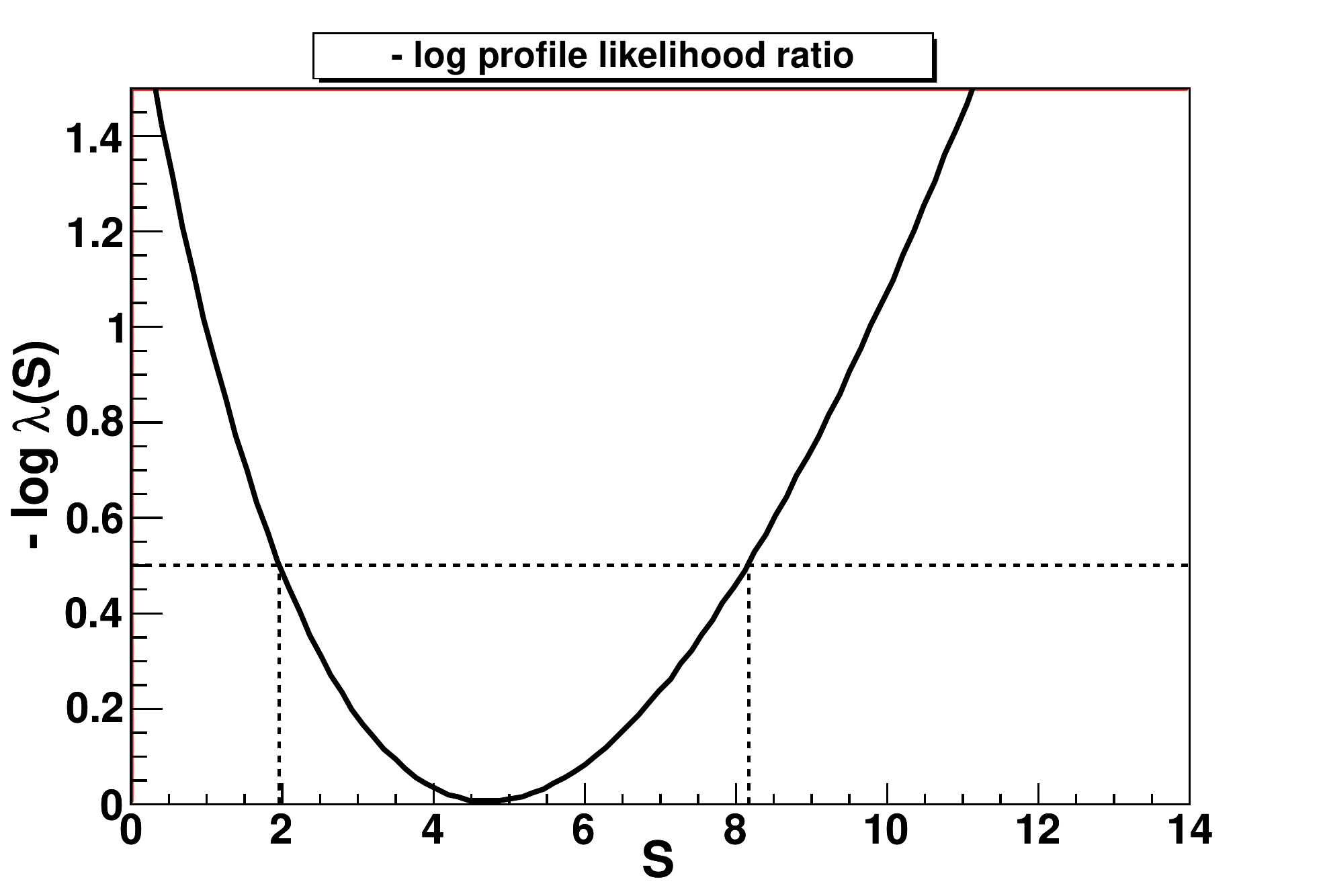}
        \caption{ Plot of the log profile likelihood curve as function
of the parameter of interest, $\theta_0\equiv S$. The $1\sigma$
interval ($68\%$ CL) is obtained from the intersect of the
$-\log\lambda$ curve with the horizontal dashed line $-\log\lambda =
0.5$. \label{likelihood_plot} }
  \end{center}
\end{figure}

Under certain regularity conditions, Wilks's theorem demonstrates that
asymptotically $-2\ln\lambda(\theta_0)$ follows a $\chi^2$
distribution. In the asymptotic limit, the likelihood ratio test statistic $\lambda(\theta_{0})$ has
a parabolic shape:
\begin{equation}
    \label{eqn:parabolicnll} -2\ln\lambda(\theta_0) = -2
(\ln{L(\theta_0)}-\ln{L(\hat{\theta}_0)}) = n_{\sigma}^{2},\
\mathrm{with}\ n_\sigma=\frac{\theta_{0}-\hat{\theta}_{0}}{\sigma},
\end{equation} where $n_\sigma$ represents the number of Gaussian standard
deviations associated with the parameter $\theta_{0}$. From this construction, it
is possible to obtain the one- or two-sided confidence intervals  (see Fig.~\ref{likelihood_plot}). Owing to the
invariance property of the likelihood ratios, it can be shown that
this approach remains valid for non parabolic log-likelihood
functions. 
This method is also known as MINOS in the physics community,
since it is implemented by the MINOS algorithm of the Minuit program.
Given the fact that asymptotically $-2\ln\lambda$ is distributed as a
$\chi^2$ variate, an hypothesis test can also be performed to
distinguish between two hypotheses characterized by different values
of $\theta_0$.

In this approach, systematic uncertainties are taken into account by
augmenting the likelihood function with terms that encode the
knowledge we have of the systematic uncertainties and the profiling is
now done over all nuisance parameters including those for the systematics. 

This likelihood-based technique for estimating an interval and
performing a hypothesis test is provided in RooStats by the {\tt
ProfileLikelihoodCalculator} class. The class implements both the {\tt
IntervalCalculator} and {\tt HypoTestCalculator} interfaces. When
estimating an interval, this calculator returns a {\tt
LikelihoodInterval} object, which, in the case of multiple parameters of
interest, represents a multi-dimensional contour. When performing a
hypothesis test, a {\tt HypoTestResult} object is returned with the
significance for the null hypothesis. Another class exists, {\tt
LikelihoodIntervalPlot}, to visualize the likelihood interval in the
case of one or two parameters of interest (as shown in
Fig.~\ref{likelihood_plot}). A newly developed class, {\tt
ProfileInspector}, allows inspection of the value of the nuisance
parameters for each value of the parameter of interest along the profile
log-likelihood curve.

\subsection{Bayesian Calculators}

Bayes theorem relates the probability (density) of a hypothesis given data to
the probability (density) of  data given a hypothesis. The inversion of the probability is 
achieved by multiplying the likelihood function (the probability of the data given an
hypothesis)
by a prior probability for the model, which is characterized by parameters of interest and,
typically, one or more nuisance parameters. This product is normalized so that
the integral of the posterior density, over all parameters, is unity.
The calculation of credible intervals, that is, Bayesian confidence intervals, 
requires the calculation of the cumulative posterior distribution.  In the Bayesian 
approach, nuisance parameters are removed by marginalization, that is, by integrating
over their
possible values. RooStats provide two different types of Bayesian
calculator, the {\tt BayesianCalculator} and {\tt
MCMCCalculator} classes, depending on the method used for performing the required 
integrations.
 
The current implementation of the {\tt BayesianCalculator} class works for a single
parameter of interest and uses numerical integration to compute
the posterior probability distribution. Various algorithms provided by ROOT for
numerical integration can be used, including those based on Monte
Carlo integration, such as implemented in the programs Vegas or Miser. 
The result of the class is
a one-dimensional interval ({\tt SimpleInterval}) obtained from the
cumulative posterior distribution.

The {\tt MCMCCalculator} uses a Markov-Chain Monte Carlo (MCMC) 
method to
perform the integration. The calculator runs the Metropolis-Hastings algorithm,
which can be configured by specifying parameters such as the number of iterations and
burn-in-steps, to
construct the Markov Chain.   Moreover, it is possible to replace the default uniform
proposal function with any other proposal function. The result of
the {\tt MCMCCalculator} is a {\tt MCMCInterval}, which can compute
the confidence interval for the desired parameter of interest from the
Markov Chain. The {\tt MCMCInterval} integrates the posterior density from
its mode downwards until the interval has a $1-\alpha$ probability content\footnote{It should be noted that these \emph{highest posterior density intervals} are not invariant
under under one-to-one reparametrisation.}. The {\tt
MCMCIntervalPlot} class can be used to visualize the interval and the
Markov chain.

Users can also input the RooStats model into the Bayesian Analysis
Toolkit (BAT) ~\cite{BAT}, a software package that implements Bayesian
methods via Markov-Chain Monte Carlo. In the latest release, BAT
provides a class, {\tt BATCalculator}, which can be used with a similar
interface to the RooStats {\tt MCMCCalculator} class. Developments
are foreseen that will further integrate BAT within
RooStats.

\subsection{Neyman Construction}

The Neyman construction is a pure frequentist method to construct an
interval at a given confidence level, $1-\alpha$, such that coverage is
guaranteed for fully-specified probability models. A detailed description of the
method is given in Ref.~\cite{James}. RooStats provides a class, {\tt
NeymanConstruction} that implements the construction. The class derives
from {\tt IntervalCalculator} and returns a {\tt
PointSetInterval}, a concrete implementation of {\tt ConfInterval}.

The Neyman construction requires the specification of an ordering rule that defines
the order in which potential observations are to be added to the interval in the space
of observations until the desired confidence level is reached. The ordering
rule is usually specified in terms of a specific test
statistic. Consequently, the RooStats class must be
configured with this information before it can produce an
interval. More information can now be provided  with the 
introduction of the interfaces {\tt
TestStatistic}, {\tt TestStatSampler}, and {\tt
SamplingDistribution}. Different test statistics are available, 
including:
\begin{itemize}
\item Simple likelihood ratio: $Q=L_1(\theta_0=1)/L_0(\theta_0=0)$,
\item Ratio of profiled likelihoods:
$Q'=L_1(\theta_0=1,\doublehat{\mathbf\theta}_{i\neq0})/L_0(\theta_0=0,\doublehat{\mathbf\theta}_{i\neq0}')$,
\item Profile likelihood ratio:
$\lambda(\theta_0)=L_1(\theta_0,\doublehat{\mathbf\theta}_{i\neq0})/L_0(\hat\theta_0,\hat{\mathbf\theta}_{i\neq0})$.
\end{itemize}

Another aspect to decide is how to sample it: assuming
asymptotic distribution, generating toy-MC experiments with nuisance
parameters fixed (used in {\tt NeymanConstruction}) or with nuisance
parameters sampled according to a prior distribution (used in {\tt
HybridCalculator}.

Common configurations, such as
the Feldman-Cousins approach --- where the ordering is based on
the profile likelihood ratio as the test statistic~\cite{FeldmanCousins}, 
can be enforced by using the {\tt
FeldmanCousins} class. A generalization of the Feldman-Cousins
procedure, when nuisance parameters are present, generating
toy Monte Carlo experiments with nuisance parameters fixed as
described in~\cite{kyle,Cranmer2003}, is also available.

The Neyman construction considers every point in the parameter space
independently. Consequently, there is no requirement that the interval be connected
nor that it have a particular structure. The result consists of a set of
scanned points labeled according to whether they are inside or outside the
interval ({\tt PointSetInterval} class). The user either specifies points in
the parameter space that are to be used to perform the construction or a
range and a number of points within the range, which will be scanned uniformly in a
grid. For each scanned point, the calculator will give the sampling
distribution of the chosen test statistic. This is typically obtained
by toy Monte Carlo sampling, but other techniques exist and can, 
in principle, be used. In particular,  newly developed code may be helpful
when testing hypotheses with very small $p$-values through the
application of importance sampling techniques.

\subsection{Hybrid Calculator}

This calculator implements a Bayesian/frequentist hybrid approach for hypothesis
testing. It consists of a frequentist toy Monte Carlo
method, as in the Neyman construction, but with a Bayesian
marginalization of nuisance parameters~\cite{CousinsHighland}. This
technique is often referred to as a "Bayesian-Frequentist Hybrid".

\begin{figure}[htb]
  \begin{center}
    \includegraphics[width=0.75\textwidth]{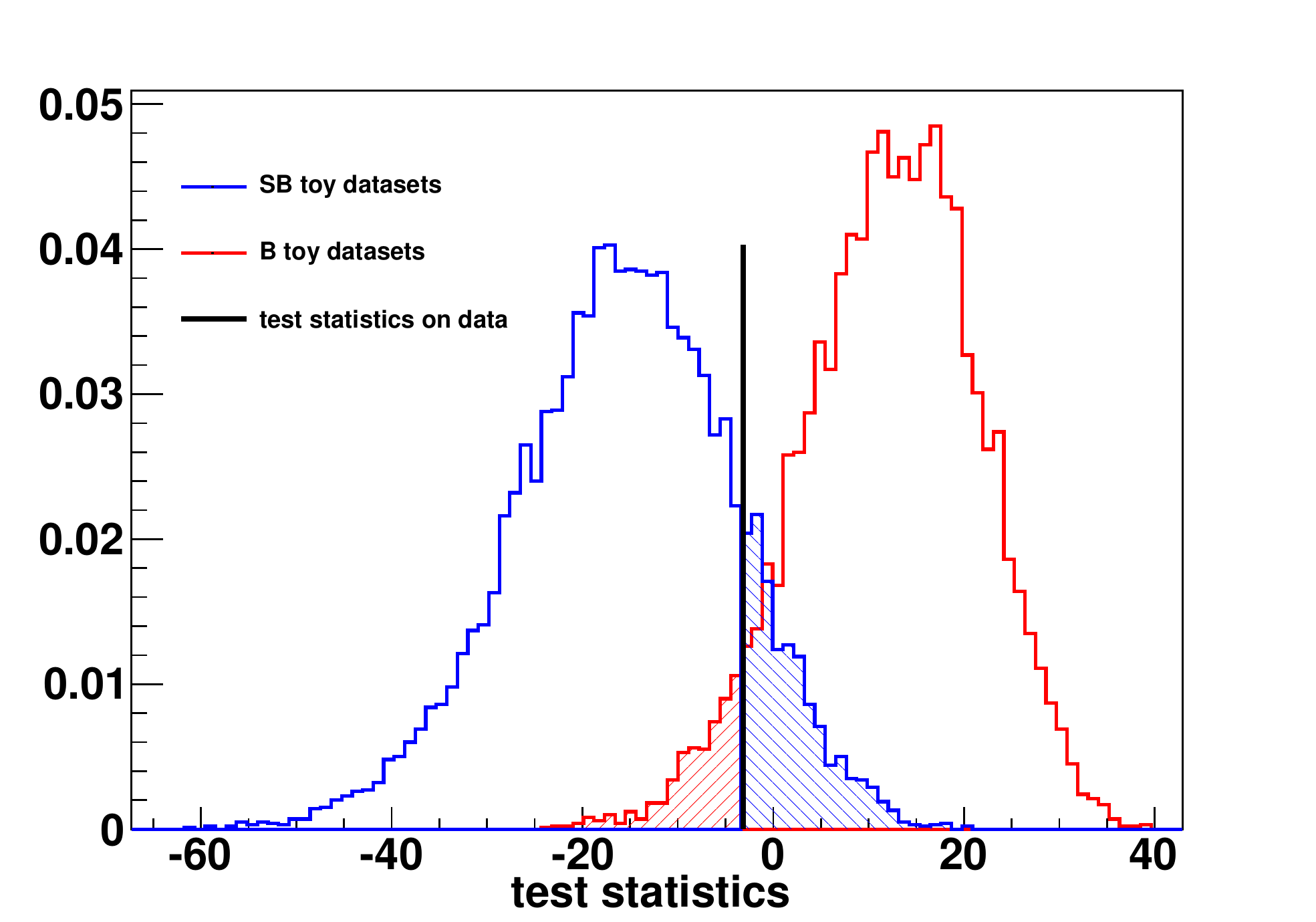}
        \caption{ Result from the hybrid calculator, the distributions
of a test statistic in the background-only (red, on the right) and
signal+background (blue, on the left) hypotheses. The vertical black
line represents the value obtained on the tested data set. The shaded
areas represent $1-CL_{b}$ (red) and $CL_{sb}$ (blue). \label{m2lnQ} }
  \end{center}
\end{figure}

For example, let us define the null hypothesis, $H_0$, to be the background-only or no signal
hypothesis, and $H_1$ to be the alternate hypothesis that a
signal is present along with background. In order to quantify the degree to which each
hypothesis is favoured or excluded by the experimental observation, one
chooses a test statistic which ranks the possible experimental
outcomes. Given the observed value of the test statistic, 
the $p$-values, $CL_{sb}\equiv p_1$ and $CL_{b}\equiv 1-p_0$,
can be computed. Since the functional forms of the test statistic
distributions are typically not known a priori, a large number of toy
Monte Carlo experiments are performed in order to approximate
these distributions. Figure~\ref{m2lnQ} provides an
example of such distributions from the two pseudo data sets
and where the observed value of the test statistic lies.

Systematics uncertainties are taken into account through
Bayesian marginalization.
For each toy Monte Carlo experiment, the 
values of the nuisance parameters are sampled from their prior distributions 
before generating the toy
sample. The net effect it to broaden the distribution of the test statistic, as
expected in the presence of systematic uncertainties, and thus degrade the
separation of the hypotheses.

This procedure is implemented in RooStats by the {\tt
HybridCalculator} class. The input to the class are the models for the two
hypotheses, the data set and, optionally, the prior distribution for the
nuisance parameters, which is sampled during the toy generation process. As
for the {\tt NeymanConstruction}, the test statistic can be freely
parameterized. The results of the {\tt HybridCalculator} consists of
the test statistic distribution for the two hypothesis, from which the
hypothesis $p$-value and associated $Z$-value can be obtained. 
Since the simulation of the distributions could be computationally
expensive, RooStats permits different
results to be merged, which makes it possible to run the
calculator in a distributed computing environment. The {\tt HybridPlot}
class provides a way of plotting the result, as shown for example in
Fig.~\ref{m2lnQ}.

By varying the parameter of interest representing the hypothesis being
tested (for example, the signal cross-section) one can obtain a
one-sided confidence interval (\eg an exclusion limit). RooStats
provides a class, {\tt HypoTestInverter}, which implements the
interface {\tt IntervalCalculator} and performs the scanning of the
hypothesis test results of the {\tt HybridCalculator} for various
values of one parameter of interest. By finding where the confidence
level curve of the result intersects the desired confidence level, an
upper limit can be derived, assuming the interval is connected. An
estimate of the computational uncertainty is also provided. Finally,
when defining exclusion limits, the condition that defines the upper
bound can be chosen: either one can use the $p$-value $p_1$ of the alternate
hypothesis (the pure-frequentist approach) or the ratio of $p$-values
$CL_s=p_1/(1-p_0)$ (modified-frequentist approach \cite{cls}).

\section{RooFit and RooStats Utilities}
\label{sec:utils}

\subsection{RooFit's Workspace}

One element of RooFit whose addition has been driven by the
development of the RooStats project (although it would still be useful
even without RooStats) is the {\tt RooWorkspace} class. It is a container for
RooFit objects that can be written to a ROOT file. When a 
RooFit object is
imported from a file (\eg, a complex PDF with multiple parameters), all the
other dependent objects are imported too. Later, it is very easy to
rebuild and initialize all the parameters, to reconstitute the original PDF,
via a single recall from the {\tt RooWorkspace} (while still permitting 
adjustments to the imported object). These features make it
possible to save the complete likelihood function, as well as the data, to a file
in a well defined fashion,  either as a technical convenience, as an
intermediate step towards the combination of the results of multiple analyses or
for the grander purpose of electronic publication of
these results. In addition, the {\tt RooWorkspace}
interfaces to a newly developed utility, {\tt RooFactoryWSTool}, which
permits the building of a large class of RooFit objects in an interpreted mode
with an intuitive syntax based on strings. Multiple dependent
parameters are also defined, created and stored in the {\tt
RooWorkspace} on-the-fly, thereby allowing, for example, the creation
of a Gaussian PDF in one line, instead of the four needed to create
one (the PDF along with its observable and two parameters) using the RooFit classes
directly. It will be discussed later how this factory tool is complemented by RooStats'
{\tt HLFactory} class.

\subsection{User-Friendly Model Specification}

Tools that simplify and automate the description of complex
models in a user-friendly way are usually referred to as model
factories. There are currently two such utilities provided within
RooStats: {\tt HLFactory} and {\tt HistFactory}. Their use is
optional. For more experienced users or in more complex cases,
direct use of the lower level RooFit classes may be preferred. 

{\tt HLFactory} is a RooStats class whose aim is to disentangle the
C++ code doing the calculations from the physics-driven and
analysis-specific description of the probability models. The later can
 be written to a single text file describing all (and only) the physics
inputs that are to be processed later in a single line of code. The
fact that {\tt HLFactory} is built as a simple wrapper around the {\tt
RooWorkspace} factory utility sidesteps the need to define yet another language
that a user would have to learn, while not restricting the application to
specific analyses since this model factory supports everything the {\tt RooWorkspace}
factory does. In addition, python-like instructions are added that
allow better structuring of the description (through includes) and along with
comments on the analysis
model. Finally (and optionally), the {\tt HLFactory} also allows the
easy combination of multiple channels to form a
combined model and combined data set.

{\tt HistFactory} is a collection of classes to handle
template histogram-based or binned analyses. It allows
such analyses to use RooStats without requiring
knowledge of the RooFit modeling language; instead, the likelihood
function and elements of the statistical analysis are specified
through an XML configuration file, which is used to 
produce the model. In this approach, the user
provides histogram templates of one observable and of 
models for different contributing samples (\eg of the signal and
background processes). Then, the normalization in terms of number of
events for each of these channels can be decomposed --- for example,
as a product of luminosity, efficiency, cross-section terms --- each
of which can be affected by systematic uncertainties. It supports Gaussian, gamma
and log-normal distributions for nuisance parameters. Finally,  histograms of variations can
 be provided that specify the related systematic changes. Multiple
channels can be given and combined and parameters
which are identical across channels can be easily identified.

\subsection{Other Utilities}

Not all utilities are listed in this document. Here we mention briefly three more:
\begin{itemize}
\item {\tt SPlot}, a class implementing a technique used to produce
weighted plots of an observable distribution in a multi-dimensional
likelihood-based analysis \cite{sPlot}.
\item {\tt RooNonCentralChiSquare}, a class in RooFit that outlines
the use of a generalization of Wilks' theorem called Wald's theorem
which states that the asymptotic distribution of the test statistic
$\lambda(\mu)$ for $\mu\neq\mu_{true}$ is a non-central $\chi^{2}$
\cite{NonCentralChiSquare},
\item {\tt BernsteinCorrection}, a class that augments the nominal
probability with a positive-defined polynomial given in the Bernstein
basis, which can be used as an approach to incorporate systematic
effects in a PDF.
\end{itemize}

\section{Statistical Combinations and Perspective}
\label{sec:theend}

The combination of results is a commonly used method for improving
sensitivities or measurements of signals. With RooStats, the
combination can be performed at the analysis level in contrast to
combinations performed at the level of published results. This means
that the global likelihood function for the ensemble of the analyses to be
combined is explicitly written and the statistical analysis is
performed on this combined likelihood. This approach
has advantages, such as being able to account for known correlations
consistently.  But, it also has its 
inconvenience, such as making the likelihood function a quite complex
object. One strong motivation for the RooStats project was to simplify
the process of combining analyses by providing a tool that allows this
to be done simply for arbitrarily complex models.

In December 2010, ATLAS and CMS created the LHC-HCG group mandated to
prepare and produce a combined Higgs result from the LHC (with
similar efforts also on-going in other analysis groups within the
collaborations). RooStats will be used for the combination and one of
the first tasks of the group has been to complement its validations
with comparison to results obtained from independent software in
specific analysis cases\footnote{For further insights on these
activities see Ref.~\cite{CramerPhystat2011}}. While the validations appear satisfactory so
far, the RooStats team will keep improving interfaces and fix
performance issues as well as develop new complementary tools based on
users' experiences and feedback.

One aspect of statistical data analysis is left open by RooStats,
namely that of the choice of statistical method. In that
respect, it allows the implementation of one recommendation of the
ATLAS and CMS statistics committees, which is that various methods be
applied and compared (although different methods are not expected to
give the same results since they have different properties and provide
answers to different questions). A more specific method and
statistical procedure to use when combining ATLAS and CMS analyses is
a topic still under discussion and one of the focuses of this PHYSTAT
conference.

\section*{Acknowledgements}

The RooStats contributors are thankful to the members of the ATLAS and
CMS statistics committees for the exchange of ideas, advice and
encouragement. I also wish to thank L. Lyons and the rest of PHYSTAT
committee for the organization of the very rich and useful conference
and the invitation to present there progress on the development of the
RooStats toolkit.

\end{document}